\documentstyle[prb,aps,multicol,epsf]{revtex}

\draft

\begin{document}

\title{Orthogonality Catastrophe in Parametric Random Matrices}

\author{Ra\'ul O. Vallejos$^{1,2}$,
        Caio H. Lewenkopf$^1$,
        and
        Yuval Gefen$^3$}

\address{$^1$
           Instituto de F\'{\i}sica,
           Universidade do Estado do Rio de Janeiro,                       \\
           R. S\~ao Francisco Xavier 524, 20559-900 Rio de Janeiro, Brazil \\
          $^2$
           Centro Brasileiro de Pesquisas F\'{\i}sicas,                    \\
           R. Dr. Xavier Sigaud 150,
           22290-180 Rio de Janeiro, Brazil                                \\
          $^3$
           Department of Condensed Matter Physics,
           The Weizmann Institute of Science,                              \\
           Rehovot 76100, Israel}

\date{\today}

\maketitle

%%%%%%%%%%%%%%%%%%%%%%%%%%%%%%%%%%%%%%%%%%%%%%%%%%%%%%%%%%%%%%%%%%%%%%%%%%%%%%

\begin{abstract}

We study the orthogonality catastrophe due to a parametric change
of the single-particle (mean field) Hamiltonian of an ergodic
system. The Hamiltonian is modeled by a suitable random matrix
ensemble. We show that the overlap between the original and the
parametrically modified many-body ground states, $S$, taken as
Slater determinants, decreases like $n^{-k x^2}$, where $n$ is the
number of electrons in the systems, $k$ is a numerical constant of
the order of one, and $x$ is the deformation measured in units of
the typical distance between anticrossings.
We show that the statistical fluctuations of $S$ are largely due
to properties of the levels near the Fermi energy.

\end{abstract}

%%%%%%%%%%%%%%%%%%%%%%%%%%%%%%%%%%%%%%%%%%%%%%%%%%%%%%%%%%%%%%%%%%%%%%%%%%%%%%


% 73.23.-b  Mesoscopic systems
% 73.23.Hk  Coulomb blockade; single-electron tunneling
% 71.10.-w  Theories and models of many electron systems
% 05.45.+b  Theory and models of chaotic systems

\begin{multicols}{2}

%-------------------------------------------------------------------------
\section{Introduction}
\label{sec:introduction}
%-------------------------------------------------------------------------

The Anderson orthogonality catastrophe (AOC), introduced by
Anderson in 1967 \cite{Anderson67}, is a fundamental effect
observed in many-body systems. The original work addressed the
ground state of a finite system consisting of $N$ non-interacting
electrons. Upon the introduction of a localized perturbation, this
ground state gets modified. Anderson has shown that the overlap
between the original and the modified $N$-electron ground states,
$\langle \Psi_N | \Psi_{N}^{'} \rangle$, is proportional to a
negative power of $N$, and vanishes in the thermodynamic limit,
hence the catastrophe. Variants of the AOC are at the basis of
some central themes in solid state physics, including the x-ray
edge singularity, zero-bias anomalies in disordered systems and
tunneling into quantum Hall systems.

The applicability of this concept and attempts to extend it to
more generic circumstances have been at the focus of attention for
more than three decades. Particularly appealing is the application
of AOC ideas to the field of mesoscopics. The study of mesoscopic
systems involves finite size systems where it is usually important
to account for the dynamics and the thermodynamics of the
electrons on a quantum-mechanical level. Important ingredients in
characterizing mesoscopic electronic systems include the strength
of the ambient disorder potential, the system's size and
shape and the strength of the electron-electron interaction.
Evidently, in most cases such systems are too complex to be
described or analyzed exactly. In such situations one needs to
resort to various approximation schemes.

Finite size (``zero-dimensional")  conductors, a.k.a. quantum dots
(QDs) have  received special attention in  recent years, partly
due to the rich physics involved, but also due to their
experimental accessibility  \cite{review,Alhassid00,Aleiner01}.
The simplest scheme to account for a QD with interacting electrons
is the {\it constant interaction} model  \cite{Ben-Jacob85}. The
latter implies that the  system Hamiltonian is given by

\begin{equation}
H=H_0 + H_{CI} \; ,
\end{equation}
where $H_0$ is the single-particle Hamiltonian and the
interaction, represented by  an infinite wavelength (zero mode),
is given by

\begin{equation}
H_{CI}=\frac{e^2}{2C}  (\hat{N} - N_0 )^2 \;.
\end{equation}
Here $C$ is the total effective capacitance of the dot, $\hat{N}$
is the particle number operator, and  $N_0$ represents a tunable
background charge (related to the gate voltage). For a complex QD
with diffusive disorder or with chaotic dynamics, the
single-particle energy spectrum and eigenfunctions of $H_0$ should
be the subject of a statistical description. It turns out that for
such systems spectral correlations within energy windows up to the
Thouless energy,\, $E_{\rm Th}$,\, are well described by the
random matrix theory (RMT), see for instance
Refs.\onlinecite{Mehta91,Guhr98}. The individual
wavefunctions, hence spectral correlations, remain unchanged as we
add/remove electrons from the system. Other than the Coulomb gap,
we do not expect any signature of the AOC.

For various reasons the constant interaction model does not
provide a satisfactory framework to account for key phenomena. The
latter include some features of the addition spectrum, Coulomb
peak-height correlations and  electron scrambling (see e.g.
Ref.\onlinecite{Alhassid00}). To improve on that model one can
employ the best effective single-particle approach, the
Hartree-Fock (HF) approximation. Expressed in the basis of the
exact eigenstates of $H_0$, denoted by  $\{\psi_{\alpha} \}$ with
the corresponding eigenenergies $\{ \varepsilon_\alpha \}$, the HF
Hamiltonian reads

\begin{equation}
H_{\rm HF}=\sum_{\alpha}  \varepsilon_\alpha \hat{n}_{\alpha}
+\frac {1}{2} \sum_{\alpha,\beta} v_{\alpha \beta \alpha \beta}
\hat{n}_{\alpha} \hat{n}_{\beta}\, .
\end{equation}
Here $v_{\alpha \beta \alpha \beta}$ are the antisymmetrized
matrix elements of the interaction and $\hat{n}_{\alpha}$ is the
number operator of the state $\psi_{\alpha}$. Note that $H_{\rm
HF}$ includes only diagonal interaction matrix elements (in the
exact eigenstate basis); for a short-ranged interaction it has
been shown  that off-diagonal matrix elements are parametrically
small. \cite{Kamenev95,Blanter97,Kurland00} It is clear that
within this approximation the effective single-particle states
(but not the  HF energies!) are independent of the occupations of
these states, and are unchanged upon the addition/removal of
electrons from the system. This is in line with the Koopmans'
picture. \cite{Koopmans}

Only when the single-particle states of the system are modified
upon the introduction of a ``perturbation" \cite{Walker99} is it
meaningful to address the question of the AOC. Here the
perturbation can be understood as varying a localized gate voltage
\cite{Vallejos98}, introducing a static impurity or adding an
electron to the $(N+1)-$st state. The latter motivates the study
of

\begin{equation}
 S=\langle \Psi_{N+1} |c^\dag_{N+1}| \Psi_{N} \rangle \,, 
\end{equation}
where $| \Psi_{N} \rangle$ is the $N$-electron system
ground-state and $c^\dag_{N+1}$ is the corresponding electron
creation operator at its first single-particle excited state.
In all these examples (including the HF scheme)
both ground states are given by Slater determinants.
Indeed the Anderson formalism deals with the overlap of two Slater
determinants. More specifically, one of the applications we have
in mind is the role of AOC in the electronic transport through
complex quantum dots. At very low temperatures their conductance
peak heights in the Coulomb blockade regime can be evaluated from
the tunneling rates to and from the corresponding
many-body ground state of the electron island
\cite{coulomb-blockade}. In the simplest approximation, these
rates are given by the overlaps between a single-particle wave
function in the dot and the channel wave functions. However, if
interactions are taken into account, even in a HF mean field
approximation, a many-body contribution to the tunneling rates has
to be considered. {\it A priori} the inclusion of an additional
electron to the island may change each individual single-particle
orbital negligibly; however, if the number of electrons in the dot
is large enough, the new many-body ground state can  be almost
orthogonal to the old one.
Some aspects of the AOC in the presence of disorder have been
previously considered, see e.g. Ref.\onlinecite{Chen92}. In the
present paper we try to circumvent the task of analyzing the AOC
in the presence of disorder potential, and replace this challenge
be studying the AOC within the framework RMT. Here the role of an
added perturbation is played by a parameter which is varied. We
end up studying the AOC in the context of parametric random
matrices \cite{Simons}. As has been argued recently
\cite{Alhassid01} this is a good model to describe electron
scrambling in quantum dots embedding the discrete electron number
in the dot in a continuous variation of a
parameter.\cite{Vallejos98}

The paper is organized as follows. In Section \ref{sec:OC} we
rederive the expression for the ground state overlaps, reviewing
the main features of the theory and introducing the notation
 employed throughout this work. Section \ref{sec:model} is
devoted to the presentation of the random matrix model used in
this study. Section \ref{sec:results} contains the main body of
our analysis, presenting our analytical and numerical results for
the {\it averaged} ground state overlaps. The distribution of
$|S|$ is the subject of Section \ref{sec:PsmallS}. Section
\ref{sec:conclusions}  contains some brief concluding remarks. We
have also included  three appendices. Appendix A quantifies the
accuracy of the ``unit volume approximation" used by Anderson in
his original work \cite{Anderson67}. In Appendix B we discuss some
implications of evaluating the ensemble-averaged value of the
Anderson integral subject to {\it grand canonical} constraints.
Finally in Appendix C we discuss the precision of the first order
perturbation theory used to evaluate Anderson's integral, defined
in Section \ref{sec:OC}.

%-------------------------------------------------------------------------
\section{Reminder on the Anderson orthogonality catastrophe: a
         parametric approach}
\label{sec:OC}
%-------------------------------------------------------------------------

Let us consider systems of electrons whose Hamiltonian, $H(X)$ is
a function of a parameter $X$. In this study the generic parameter
$X$ can be either continuous or discrete. In the case of modelling
the change in an external magnetic field or in the electrostatic
potential, $X$ is continuous. One can also imagine $X$
as modelling the change in the {\it effective} single-particle
potential due to the sequential addition of  electrons to the
system, in which case $X$ is discrete (in the latter case it is
necessary first  to subtract all systematic  changes of the
single-particle spectrum.) It has been shown recently
\cite{Alhassid2001b} that scrambling due to the sequential
addition of electrons to the QD can be embedded in a parametric RM
process only if interaction matrix elements due to the
accumulation of surface charge are taken into account. The subject
of the present paper is the overlap $S$ between ground states of
systems corresponding to $H(X)$ and $H(X + \delta X)$. In
particular, this section is devoted to the presentation of
different levels of approximation for $\log |S|$.

Since we are dealing with non-interacting electrons, the ground
state wave functions are Slater determinants. The overlap $S$ can
then  be expressed in terms of the occupied single-particle wave
functions overlap. Let us denote by $\psi_k(X)$ and
$\psi_k(X+\delta X)$ the (single-particle)  eigenstates of $H(X)$
and $H(X + \delta X)$ respectively. We define the unitary overlap
matrix $A$ as
\begin{equation}
\label{eq:defA}
A_{ij} = \langle \psi_i(X) | \psi_j(X+\delta X) \rangle \; .
\end{equation}
Thus the overlap of the ground states $S$ can be written as
\begin{equation}
      S =  \det A^{\rm oo} \; ,
\end{equation}
where the superscript ``o" stands for occupied states. Accordingly
$A^{\rm oo}$ corresponds to the subspace of occupied states of the
matrix $A$ defined in Eq.\ (\ref{eq:defA}).

The catastrophe is manifest by the fast suppression of the overlap $S$
as the number of fermions $n$ in the system is increased for a fixed
perturbation strength.
Anderson \cite{Anderson67} singled out two basic reasons that make
the absolute value of the overlap $S$ smaller than one:
The rows of $A^{\rm oo}$ have norms smaller than unit and they do not
form an orthogonal set.
It is convenient to separate these two contributions by introducing a
new matrix $\widetilde{A^{\rm oo}}$ with normalized rows, defined as
\begin{equation}
     \left( \widetilde{A^{\rm oo}} \right)_{ij} =
     \left( A^{\rm oo} \right)_{ij}/{\cal N}_i \; ,
\end{equation}
where the normalization factor ${\cal N}_i$ reads
\begin{equation}
{\cal N}_i =
\sqrt{ \sum_{j=1}^{n}\left[ \left( A^{\rm oo}
                      \right)_{ij} \right]^2 }\; .
\end{equation}
$|S|$ is then rewritten as
\begin{equation}
     |S|=
         |\det \widetilde{A^{\rm oo}}| \times
          \prod_{i=1}^{n} {\cal N}_i   \; .
\end{equation}
The absolute value of $\det \widetilde{A^{\rm oo}}$ can be
geometrically interpreted as the volume of a $n$-dimensional
parallelepiped defined by a set of non-orthogonal vectors of unit
length. In Ref. \onlinecite{Anderson67} it is claimed that $ |\det
\widetilde{A^{\rm oo}}| $ does not depend on $n$ and it is close
to one. In Appendix \ref{app:perturbation} we show
perturbatively   that indeed the most important contributions to
$|S|$ come from $\prod {\cal N}_i$, the deviations from the ``unit
volume approximation" being of higher order in $\delta X$. Aiming
at obtaining  an upper bound of the overlap $S$ we shall neglect
below the contribution of $|\det \widetilde{A^{\rm oo}}|$ to
$|S|$. We thus restrict our analysis to the product of
normalization factors. We can therefore  write
\begin{equation}
\label{eq:firstapp}
     |S| <  \prod_{i=1}^{n} {\cal N}_i
      \equiv  \prod_{i=1}^{n} ( 1- P_i )^{1/2}  \; ,
\end{equation}
where we have introduced
\begin{equation}
\label{eq:defPi}
     P_i = \sum_{j=n+1}^{\infty}
                |\langle \psi_i(X) | \psi_j(X+\delta X) \rangle|^2
                             \;.
\end{equation}
$P_i$ measures the probability of the state $i$, whose energy
$\varepsilon_i$ is smaller then the Fermi energy, to be spread
over the $j$-components  lying above the Fermi surface once the
Hamiltonian is changed. The product of the normalization
factors, Eq.\ (\ref{eq:firstapp}), is the first level of approximation
to the overlap determinant  considered here. For the sake of
convenience, hereafter we shall deal with $\log |S|$ instead
of $|S|$ itself, and define
\begin{equation}
\label{eq:defL1}
    I_{\mbox{\scriptsize norm}} \equiv
          \frac{1}{2} \sum_{i=1}^{n} \log (1-P_i) > \log|S| \;.
\end{equation}
Eq.~(\ref{eq:defL1}) is still not suitable for an insightful
analysis. For this purpose a higher level of approximation for $S$
was introduced \cite{Anderson67}. For small $\delta X$ the
``probabilities'' $P_i$ will also be small. Thus, by
expanding $I_{\mbox{\scriptsize norm}}$ in a power series
\begin{equation}
    I_{\mbox{\scriptsize norm}}  = \frac{1}{2} \sum_{i=1}^{n}(-P_i -
                         \frac{P_i^2}{2}- \ldots)
\end{equation}
and retaining only its first expansion term one writes
\begin{equation}
  I \equiv -\frac{1}{2} \sum_{i=1}^{n} P_i > I_{\mbox{\scriptsize norm}} \;.
\end{equation}
By explicitly expressing $I$ in terms of the single-particle
overlaps, one arrives at
\begin{equation}
\label{eq:defL2}
    I =  -\frac{1}{2} \sum_{i=1}^{n} \sum_{j=n+1}^\infty
  |\langle \psi_i(X) | \psi_j(X+\delta X) \rangle|^2
  \; .
\end{equation}
This is the well known {\it Anderson integral} term. The
approximations $I_{\mbox{\scriptsize norm}}$ and $I$ for $\log
|S|$ will be analyzed at length in Section \ref{sec:results} where
we present numerical and analytical results for the random matrix
model introduced in the forthcoming section.

%-------------------------------------------------------------------------
\section{The random matrix model}
\label{sec:model}
%-------------------------------------------------------------------------

In this Section we present a model for the statistical study of
the Anderson orthogonality catastrophe in ballistic ergodic
systems. The setting is the same as in the foregoing Section. We
model the single-particle Hamiltonian $H$ which depends on a
parameter $X$  by a suitable ensemble of random matrices. In one
of its simplest forms \cite{Wilkinson89}, such model is
realized by
\begin{equation}
\label{eq:model}
      H(X) = H_0 \cos X  + U \sin X \; ,
\end{equation}
where $H_0$ and $U$ are two independent matrices of dimension $N
\times N$, both members of the same Gaussian Ensemble.
In this study, we restrict our analysis to the Gaussian Orthogonal
Ensemble ($\beta = 1$) and to the Gaussian Unitary Ensemble
($\beta=2$). Whereas the orthogonal ensemble is used to model
systems presenting time-reversal symmetry, the unitary ensemble
models systems where the latter symmetry is absent. Thus the
distribution of $H_0$ and $U$ is \cite{Mehta91,Guhr98}
\begin{equation}
\label{eq:P(H)}
  P(H_0, U) = C_{\beta N} \exp \left\{ -\frac{2N}{\beta \lambda^2}
            (\mbox{tr}\, H_0 H_0^\dagger +
         \mbox{tr}\,  U U^\dagger)\right\} \;,
\end{equation}
where $C_{\beta N}$ is a normalization constant.
As a consequence of this choice, the matrix elements of $H(X)$
are also Gaussian distributed with zero mean and variance parameterized as
\begin{equation}
\label{eq:variance}
     \langle H_{ij}^*(X) H_{i^\prime j^\prime}(X) \rangle  =
     (\delta_{i i^\prime} \delta_{j j^\prime} + \delta_{\beta 1}
      \delta_{i j^\prime} \delta_{j i^\prime}) \, \frac{\lambda^2}{N}
\end{equation}
independent of $X$. Here the symbols $\langle \cdots \rangle$
stand for ensemble averaging.

For any arbitrary value of $X$, the resulting mean level density is given
by the well-known Wigner semicircle formula \cite{Mehta91}
\begin{equation}
\label{eq:semicircle}
   \rho(\varepsilon) =  \frac{N}{\pi \lambda}
      \sqrt{1 -  \left(\frac{\varepsilon}{2\lambda}\right)^2 }   \; ,
\end{equation}
where $\varepsilon$ is the single-particle energy. Evidently, the
mean level spacing at the center of the spectrum ($\varepsilon=0$)
is $\Delta=\pi \lambda/N$. Accordingly, the average band width is
$\varepsilon_{\rm max} - \varepsilon_{\rm min} = 4\lambda$ (this
is in line  with standard solid state picture whereby the band
width is fixed, independent of the system's size, while the mean
level spacing scales  as $N^{-1}$).

The typical scale for the parametric variation $X^\star$
representing the average distance between anticrossings, can in
general be characterized by the level velocity standard deviation
\cite{Szafer93,Simons93}
\begin{equation}
  \frac{\Delta}{X^\star} = \sqrt{
  \left \langle \left( \frac{d \varepsilon_\mu }{d X}\right)^2
                                                  \right \rangle -
  \left \langle  \frac{d \varepsilon_\mu }{d X} \right \rangle^2  } \; ,
\end{equation}
where $\varepsilon_{\mu}(X)$ is an energy level of $H(X)$ close to
the center of the band.
In the model defined by Eqs.~(\ref{eq:model}) and (\ref{eq:P(H)})
the average level velocity
is zero and
\begin{equation}
\label{eq:defXstar}
  X^\star = \frac{\Delta}{\sqrt{2/(\beta N)} \, \lambda} =
    \pi \sqrt{\frac{\beta}{2N}}  \; .
\end{equation}
Now it is possible to quantify the effect of $\delta X$ in
the single-particle spectrum by the scaled parameter $x \equiv
\delta X/ X^\star$.

It remains to specify the many-body part of the model. The ground
state configuration is generated by populating the single-particle
levels with $n=N/2$ fermions (the spin degree of
freedom is not considered here). For a given  $X$, the ground
state is the Slater determinant made up of the  lowest
$N/2$ eigenstates of $H(X)$. Now the single particle
Hilbert space is finite and of size $N$, and the number of
particles is $N/2$. The Fermi level, $\varepsilon_F =
(\varepsilon_{N/2+1} + \varepsilon_{N/2})/2$, lies on average at
$\varepsilon_F = 0$. It is important to stress that the sums in
Eqs.~(\ref{eq:defPi}) and (\ref{eq:defL2}) have to be modified
accordingly.

Numerical simulations of the  model presented above are
straightforward to implement. We generate a pair of matrices $H_0$
and $U$ of dimension $N$ whose elements are Gaussian distributed
with zero mean and variance given by Eq.~(\ref{eq:variance}).
Fixing $X$ and $x$ we calculate the eigenvalues and eigenfunctions
of the corresponding $H(X)$ and $H(X+\delta X)$. Having computed
the complete set of eigenfunctions, we calculate the overlap
matrix $A^{\rm oo}$ and its determinant, $S_N(X, x)$. In order to
keep the notation simple, we shall write down the
arguments of $S$ explicitly only  when necessary.

\begin{figure}
\setlength{\unitlength}{1mm}
\begin{picture}(100,105)
\put(20,  25){\epsfxsize=67mm\epsfbox[200 208 600 600]{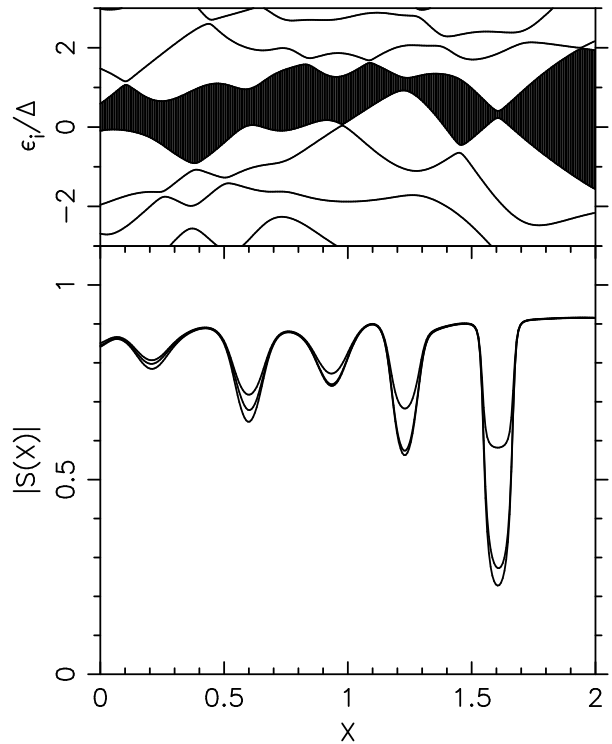}}
\end{picture}
\caption{Top panel: typical energy levels as a function of $X$.
         The filled region is the gap between the last occupied and the
         first empty single-particle levels. Bottom: the three curves
         represent the exact overlap $S$, $\exp(I_{\mbox{\scriptsize norm}})$,
         and $\exp(I)$ (see text for definitions) as a function of $X$.
         They respect the ordering  $S< \exp(I_{\mbox{\scriptsize norm}}) <
         \exp(I)$.
         The parametrical distance is set to $x=0.4$ ($\delta X \approx
         0.13$) and the dimension of the single particle  Hilbert
         space is $N=50$. }
\label{fig:typical}
\end{figure}

As a guide for the discussions to come  let us consider a
representative member of the ensemble defined by Eq.\
(\ref{eq:P(H)}) for $\beta =1$ and $N=50$. In
Fig.~\ref{fig:typical} we show the parametric spectrum as a
function of $X$ around $\varepsilon=0$ (top panel) and the
corresponding ground state overlaps $|S(X)|$ for $x=0.4$ (bottom
panel),  both obtained by a direct numerical procedure. The most
striking feature of this plot is that large fluctuations in $|S|$
are closely correlated with the occurrence of narrow avoided
crossings at the Fermi surface. In other words, to each narrow gap
(avoided crossing) there corresponds a small overlap $|S|$.

\noindent
The figure also suggests that the approximation schemes presented in
Section \ref{sec:OC} fail in the vicinity of narrow gaps. These
observations indicate that an accurate analytical estimate of the
average $|S|$ and its variance requires  a good handle of
the ``two-level problem" of narrow avoided crossings. Had we
considered integrable (or mixed) systems, this issue would become
even more important due to the absence (or suppression) of level
repulsion.

%-------------------------------------------------------------------------
\section{Average ground states overlap}
\label{sec:results}
%-------------------------------------------------------------------------

As we shall argue in the next section, the distribution of $S$
can be anomalously broad.
As a first step though, we shall consider in this section
{\it ensemble averaged} quantities.
When the distribution is narrow, as is often the case with
thermodynamic quantities, one may interchange
\begin{equation}
\label{eq:interchange}
\langle \log |S| \rangle \approx \log
\langle |S| \rangle \,.
\end{equation}
Evidently, when in doubt, the
meaningful quantity is the r.h.s. of  the equation above.
Technically, though, the more accessible quantity is the l.h.s. of
Eq.\ (\ref{eq:interchange}). This section is devoted to the
statistical study of upper bounds for $\langle \log |S| \rangle$
within the approximation levels presented in Section \ref{sec:OC}
and the random matrix model described in Section \ref{sec:model}.
Our analytical results  are complemented with numerical
simulations.

In Eq.~(\ref{eq:defL2}) the ground states overlap can be easily
evaluated by considering the single-particle overlaps in first order
perturbation theory, namely
\begin{equation}
\label{eq:fopt}
\langle \psi_i(X) | \psi_j(X+\delta X) \rangle  \approx
\frac{ \delta H_{ij} }{ \varepsilon_j - \varepsilon_i } \; .
\end{equation}
Here $\delta H \approx (- H_0 \sin X + U \cos X ) \, \delta X$
and $\{\varepsilon_j\}$ are the eigenvalues of $H(X)$.
The matrix $\delta H$ and the set of eigenvalues $\{\varepsilon_j\}$
are statistically independent due to the invariance of the considered
ensembles under orthogonal $(\beta=1$) or unitary $(\beta=2)$
transformations.
Recalling Eq.~(\ref{eq:defL2}) $\langle\log |S|\rangle$ now reads
\begin{equation}
\label{eq:perturba}
    \big\langle \log |S| \big\rangle
     < \langle \widetilde{I} \rangle  \equiv
   - \frac{1}{2} \sum_{i=1}^{N/2} \sum_{j=N/2+1}^{N}
   \left \langle
   \frac{ |\delta H_{ij}|^2 }{( \varepsilon_j - \varepsilon_i)^2 }
   \right \rangle   \;.
\end{equation}
In Appendix \ref{app:perturbation} it is shown that the
corrections responsible for the inequality in
Eq.~(\ref{eq:perturba})  are of fourth order in a perturbation
expansion. Note that only off-diagonal $\delta H_{ij}$ matrix
elements contribute  in Eq.~(\ref{eq:fopt}) since the $i$-states
lie below the Fermi surface and the label $j$ corresponds to
states above it. The ensemble average over $|\delta H_{ij}|^2$,
defined in Eq.~(\ref{eq:variance}), yields
\begin{equation}
\label{eq:averageoverdH}
     \langle \widetilde{I} \rangle
     = - \frac{1}{2} \sum_{i=1}^{N/2} \sum_{j=N/2+1}^{N}
          \frac{\beta}{2} x^2
          \left \langle
      \frac{\Delta^2} {\left( \varepsilon_j - \varepsilon_i \right)^2}
      \right \rangle  \; ,
\end{equation}
where we have used $X^\star$ of Eq.~(\ref{eq:defXstar}) to express
$\big\langle \log |S| \big\rangle$ in terms of the dimensionless
variable $x = \delta X/X^\star$.

 A satisfactory accuracy of the approximation introduced by Eq.\
(\ref{eq:fopt}) requires $\delta H_{ij}/(\varepsilon_j -
\varepsilon_i)$ to be small. Such a condition is translated to
$\langle \widetilde{I} \rangle$ by examining separately $x^2$ and
$\langle \Delta^2/( \varepsilon_j - \varepsilon_i )^2\rangle$.
Viewed from the perspective of the parametric framework, the
approximation is under control since the physical situations
in mind, see Section \ref{sec:introduction}, call for $x\ll
1$, or at least for $x <1$ \cite{review}. By contrast,
the average over $\Delta^2/( \varepsilon_j - \varepsilon_i )^2$
requires a careful discussion.

The occurrence of small gaps at the Fermi surface, $\varepsilon_j
- \varepsilon_i \ll \Delta$ for $j-i=1$, causes the breakdown of
the perturbation approach employed in Eq.~(\ref{eq:fopt}), to
calculate $\langle \psi_i(X) | \psi_{i+1}(X+\delta X)
\rangle$. In that case even a small variation of $x \ll 1$
may give rise to a significant mixing of an originally occupied
level (the last one) and an empty one (the lowest original vacant
level). Such a situation calls for a non-perturbative solution.
Indeed it is  tempting  to justify the approximation used
in Eq.~(\ref{eq:fopt}) invoking the presence of level repulsion:
the large errors introduced in the calculation of $|S|$ at small
gaps are minimized by the rareness of such events. Unfortunately,
such a scheme is doomed to fail due to the following
reason: For neighboring levels
\begin{equation}
\label{eq:naive}
 \left \langle
    \frac{\Delta^2} {\left( \varepsilon_{i+1} - \varepsilon_i \right)^2}
 \right \rangle  =
     \int_0^\infty \! ds\, s^{-2} P_\beta (s) \;,
\end{equation}
where $P_\beta(s)$ is the nearest neighbor spacing distribution
\cite{Mehta91}. Since, $P_\beta(s) \sim s^\beta$ for $s \ll 1$ the
average on the l.h.s. of Eq.~(\ref{eq:naive}) diverges for the
orthogonal symmetry (although not for the unitary case).  This
motivates the employment of a non-perturbative approach to account
for the effect of the ``crust levels" (near the Fermi energy) on
the ensemble-averaged Anderson integral. With this proviso in mind
it should be also realized that $j-i=1$ corresponds to a single
term in the double sum of Eq.~(\ref{eq:averageoverdH}). As
concluded from Fig.~\ref{fig:typical} this term has a {\it major
contribution to the large fluctuations} in $|S|$ and will be the
subject of analysis of Section \ref{sec:PsmallS}.
By now it is only necessary to anticipate that the occurrence of
narrow gaps contributes to  $\langle \log|S| \rangle$ with an
additional factor which depends on $x$ and on the size of the
narrow gap. Note that the procedure presented in
Eq.~(\ref{eq:naive}) represents averaging under {\it canonical}
conditions, namely  averaging over systems with a given particle
number $N$ \cite{Shklovskii82} or averaging over both impurity
realizations and $N$. Such procedures are referred to as strong
and weak canonical averaging, cf. Ref.\onlinecite{Kamenev94}.
There are experimental setups where it is the chemical potential,
$\mu$, which is the controlled parameter. In such circumstances it
is more appropriate to employ a grand-canonical averaging
procedure. This is briefly discussed in Appendix \ref{app:GC}.

As the value of $j-i$ becomes larger the perturbative approach
works increasingly better. Moreover, at the same time the
fluctuations of $\varepsilon_j - \varepsilon_i$ relative to $(j -
i)\Delta$ decrease as a consequence of  the spectral rigidity.
These matters are discussed in Appendix \ref{app:FOPT}. In this
regime we do not expect to introduce a large error (independent of
$N$) replacing the spectrum of $H(X)$ by its average spectrum,
that is
\begin{equation}
           \left \langle
       \frac{1} {\left( \varepsilon_j - \varepsilon_i \right)^2}
       \right \rangle
   \approx
       \frac{1}
       { \left( \langle \varepsilon_j \rangle -
                \langle \varepsilon_i \rangle \right)^2 } \; .
\end{equation}
The latter approximation allows us write
\begin{equation}
     \langle \widetilde{I} \rangle =
     - \frac{\beta}{4} x^2
       \int^{-\Delta/2}_{-2\lambda} d\varepsilon_i
       \int_{ \Delta/2}^{ 2\lambda} d\varepsilon_j \,
       \frac{\rho(\varepsilon_i) \rho(\varepsilon_j)}
       {(\varepsilon_j - \varepsilon_i)^2} \; ,
\end{equation}
with the mean level density $\rho(\varepsilon)$ given by the
Wigner semicircle law, Eq.~(\ref{eq:semicircle}). Changing
variables and defining $\alpha \equiv \pi/(4N)$ we arrive at
\begin{equation}
\label{eq:Itilde-quasela}
     \langle \widetilde{I} \rangle =
   -  \frac{\beta}{4} x^2
      \int_{ \alpha}^{ 1} \! du
      \int_{ \alpha}^{ 1} \! dv \,
      \frac{ \sqrt{1-u^2} \sqrt{1-v^2}}
           { (u+v)^2}             \; .
\end{equation}
We can now isolate the singularity at $\alpha = 0$ ($N$ large) and
rewrite the integral on the r.h.s. of
Eq.~(\ref{eq:Itilde-quasela})
\begin{equation}
\label{eq:isola}
     \int_{ \alpha}^{ 1} \! du
      \int_{ \alpha}^{ 1} \! dv \,
      \left[\frac{ 1}{ (u+v)^2}    +
      \frac{ \sqrt{1-u^2} \sqrt{1-v^2} - 1}
           { (u+v)^2} \right]                         \; .
\end{equation}
While it is the first term under the integral in  Eq.\
(\ref{eq:isola}) which is ultimately responsible for the
catastrophe, the second one gives only a constant as $\alpha
\rightarrow 0$. The final result is therefore
\begin{equation}
\label{eq:teo}
 \langle \widetilde{I} \rangle
     = - \frac{\beta}{4} x^2
                  \left( \log N + C \right) \; ,
\end{equation}
where $C = -\log(\pi/2) - \pi^2/8$. The latter constant
has to be taken with caution due to the approximations made.

Summarizing our results, we have several levels of approximation
to the exact value of $\langle \log |S| \rangle$. The product of
the normalization factors, which neglects corrections to the
volume of the parallelepiped, assumed to be unity, yields
\begin{equation}
\label{eq:s_average}
\langle \log |S| \rangle \approx \langle
I_{\mbox{\scriptsize norm}} \rangle \equiv \frac{1}{2} \sum_{i=1}^{N/2}
\langle \log (1-P_i) \rangle \; . 
\end{equation}
By keeping only the first term of the expansion of the logarithm
in $I_{\mbox{\scriptsize norm}}$ we have
\begin{equation}
\label{eq:s_average2}
\langle \log |S| \rangle \approx \langle I
\rangle \equiv -\frac{1}{2} \sum_{i=1}^{N/2} \langle P_i \rangle \; .
\end{equation}
By calculating $I$ using first order perturbation theory and a
smoothed spectrum, we obtain the analytical estimate
$\langle \widetilde{I} \rangle$, given by Eq.~(\ref{eq:teo}).

For small values of $x$ these quantities are close to each other.
They are ordered as follows:
\begin{equation}
\langle \log |S| \rangle  <  \langle I_{\mbox{\scriptsize norm}} \rangle  <
  \langle I \rangle < \langle \widetilde{I} \rangle       \; .
\end{equation}

We turn now to our numerical analysis. For different values of $x$
and $N$ we have evaluated $\langle \log |S| \rangle, \langle
I_{\mbox{\scriptsize norm}} \rangle, \langle I \rangle$, and
$\langle \widetilde{I} \rangle$ ensemble averaging over $M$
realizations of $H(X)$ which is defined in Section
\ref{sec:model}. Each simulation is performed at the cost of the
order of $M \times (\beta \times N)^3$ operations, imposing a
computational constraint on the procedure.

\begin{figure}
\setlength{\unitlength}{1mm}
\begin{picture}(100,115)  
\put(26,  25){\epsfxsize=67mm\epsfbox[200 208 600 600]{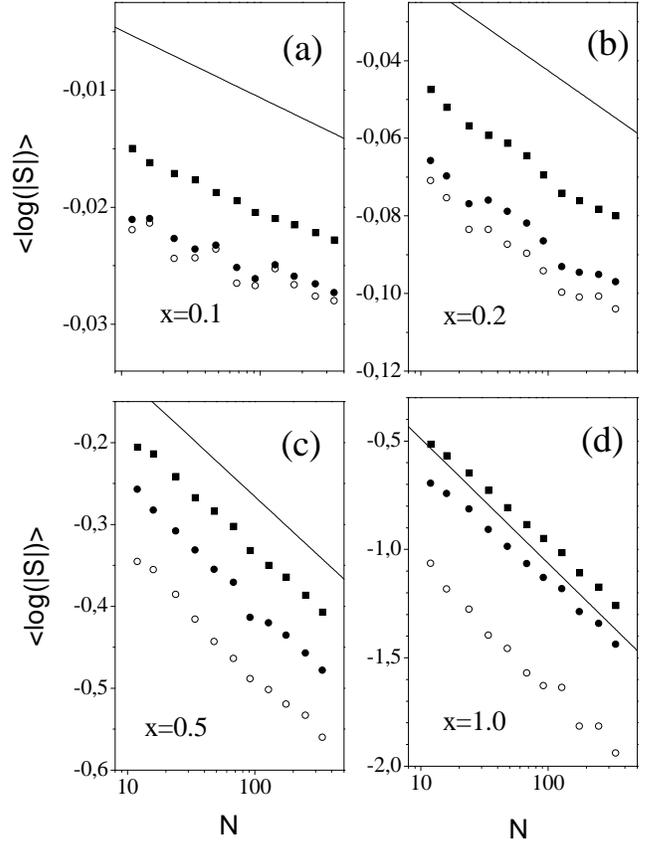}}
\end{picture}
\caption{The average $\log |S(x)|$ as a function of $N$.
         The number of
         realizations for each $N$  is $M=10^4$
         and (a) $x=0.1$, (b) $x=0.2$, (c) $x=0.5$, and (d) $x=1.0$.
         Open dots stand for the exact $\langle \log |S| \rangle$,
         filled dots for $\langle I_{\mbox{\scriptsize norm}}\rangle$,
         squares for $\langle I \rangle$, whereas the solid lines
         represent $\langle \widetilde{I} \rangle$.}
\label{fig:canon-average}
\end{figure}

In Fig.~\ref{fig:canon-average} we show a comparison between the
different approximations for $\langle\log|S|\rangle$ as a function
of $N$ for $\beta=1$. We chose four representative values of $x$
and fixed $M=10^4$.  We observe that in all approximation schemes
$\langle \log |S| \rangle$ displays the slope predicted by
Eq.~(\ref{eq:teo}). In view of the large sample sizes, the 
fluctuations in the mean values indicate the corresponding 
distributions are characterized by large standard deviations, as
we discuss in the following section. As analyzed in Appendix 
\ref{app:FOPT} the
first order perturbation theory estimate breaks down when used for
levels in the vicinity of the Fermi surface  and/or for not
sufficiently small values of  $x$. As the latter is increased we
even expect deviations from the predicted slopes for $\langle I
\rangle$ versus $\log N$. In Fig.~\ref{fig:canon-average} such
discrepancies are only noticeable for $x=1$ at $N \le 50$. We
conclude that the power law suppression of $|S|$ as a function of
$N$ is very robust. The discussion presented in Appendix
\ref{app:FOPT} suggests why our theoretical estimate does not
always serve as an upper bound for all our  approximation schemes.
This is done by noting two facts. (i) For small values of
$x$ the perturbation approach underestimates $\langle |
A_{N/2,N/2+j} |^2\rangle$, hence overestimates the overlap. In
such cases we are guaranteed an upper bound. (ii) For larger
values of $x$ the situation is quite the opposite. When this
happens the constant $C$ does not provide any longer  an
upper bound. Results for the  $\beta = 2$ case behave in the
very same way as for $\beta = 1$, with slopes following Eq.\
(\ref{eq:teo}). (We have thus decided to omit a
corresponding figure for the unitary case).

Figure \ref{fig:canon-average} also confirms Anderson's claim that
the ``unit volume" corrections to $\langle \log |S|\rangle$ are
small, at least for $x\ll 1$, as indeed is seen in panels (a) and
(b). While within our statistical precision we do not
observe any $N$ dependence for the unit volume corrections, the
latter cannot be ruled out.

%-------------------------------------------------------------------------
\section{Distribution of overlaps and their large fluctuations}
\label{sec:PsmallS}
%-------------------------------------------------------------------------

This section is devoted to the analysis of the distribution of the
overlaps between ground states, $P(|S|)$. We show that the large
fluctuations of $S(X)$ occurring in the vicinity of small gaps at
the Fermi surface are well described by a $2\times 2$ model,
otherwise $\exp(\langle \widetilde{I} \rangle)$ provides
a satisfactory estimate for  $|S|$.

\begin{figure}
\setlength{\unitlength}{1mm}
\begin{picture}(70,70)
\put( 20,  25){\epsfxsize=65mm\epsfbox[200 208 600 600]{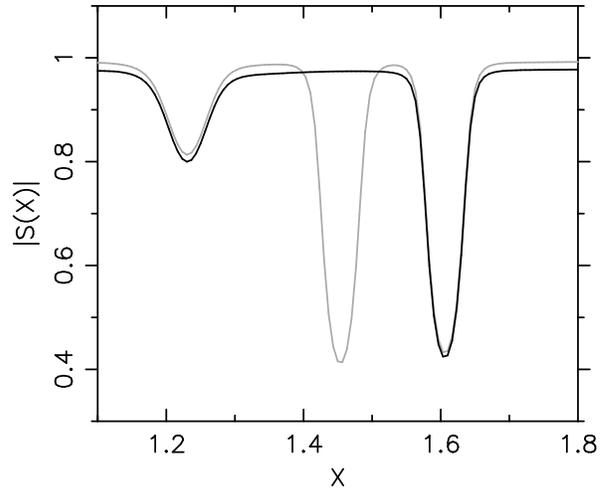}}
\end{picture}
\caption{Overlaps of Slater determinants
for $x=0.2$ as a function of $X$ for $N=50$.
The exact determinant is shown as a black line.
The grey line corresponds
to the approximation
$|S| \approx |\langle \psi_{N/2}(X) | \psi_{N/2}(X+\delta X) \rangle| $.}
\label{fig:lastlevel}
\end{figure}

Let us start quantifying  the influence of the narrow avoided
crossings on the overlap determinant.
In the proximity of a narrow gap the single-particle overlap
$\langle \psi_{N/2}(X) | \psi_{N/2}(X+\delta X) \rangle $ will be small
and rapidly varying with $X$.
Consequently, as it was qualitatively established in Section \ref{sec:OC},
it will dominate the fluctuations of the single-particle overlap matrix
determinant.
In this situation we expect that the behavior of
$S$ will be captured by the approximation
\begin{equation}
\label{eq:factorization}
 |S_N(X,\delta X)| \approx K_N(X,\delta X)
 |\langle \psi_{N/2}(X) | \psi_{N/2}(X+\delta X) \rangle |  \;.
\end{equation}
This approximation is just the first term of the expansion of
the determinant of $A^{\rm oo}$ along its $N/2$-th row.
In other words, we calculate $ |S|$ as if all fluctuations
are due to the interaction of the highest occupied single-particle
level with the lowest empty one.
The factor $K$ accounts for the contribution of all remaining
states and will be approximately constant:
 \begin{equation}
 K_N(X,\delta X) \approx \big\langle |S_{N-1}(X,\delta X)| \big\rangle \;.
\end{equation}
The results presented in Fig.~\ref{fig:lastlevel} confirm that
this approximation works impressively well  {\it in the
vicinity of  narrow gaps} and for small values of $x$. The system
is the same as that in Fig.~\ref{fig:typical},  but smaller
deformation is considered, $x=0.2$, as well as  a smaller
parameter range for $X$. The latter contains, in the
present example, two narrow gaps (cf. Fig.\ \ref{fig:typical}).
The small shift between the two curves can be attributed to
the $K$ term.  The proposed approximation gives rise to spurious
peaks whenever an avoided crossing (narrow gap) between the
occupied levels $N/2$ and $N/2-1$ is encountered. Such cases are
beyond the scope of our approximation: the factorization put
forward by Eq.\ (\ref{eq:factorization}) is no longer valid, and
the  gap at the Fermi energy  is certainly not small. The spurious
peak at $X\approx 1.45$ in Fig.\ \ref{fig:lastlevel}
represents  such a situation, as  can be verified from
Fig.\ \ref{fig:typical}.
We have thus  demonstrated heuristically that  small Anderson
overlaps arise due  avoided crossings at the Fermi level,
involving the highest occupied and the lowest unoccupied levels.
This observation leads us to conjecture that
\begin{equation}
\label{eq:2x2a}
 P(|S|) \approx P(K \cdot |\widetilde{S}|)
   \qquad \mbox{for} \qquad |S|\ll 1\;,
\end{equation}
where $\widetilde{S} = \langle \psi_{N/2}(X) | \psi_{N/2}(X+\delta
X) \rangle $. The derivation of an expression for
$P(|\widetilde{S}|)$, which calls for a non-perturbative approach,
is discussed now. Due to level repulsion two levels rarely come
very close to each other. When this happens, it is possible to
effectively model a narrow avoided crossing by a $2 \times 2$
matrix, since it is quite unlikely to have yet another level very
close by. This model also rids of  the undesired
occurrence of spurious peaks as the one shown by Fig.\
\ref{fig:lastlevel}: they cancel between  the factor $K$ and
$\widetilde{S}$. While this is quite a simple model,  the
calculation of the distribution $P(|\widetilde{S}|)$ involves some
cumbersome  multidimensional integrals \cite{Yuval}.

For the orthogonal case, $\beta=1$, the integrations can be
simplified through a geometrical construct, which makes it
possible to obtain $P(|\widetilde{S}|)$ analytically within a 2
$\times$ 2 parametric random matrix model. Let us write our model
Hamiltonian, Eq.~(\ref{eq:model}), $H(X) = H_0 \cos X + U \sin X$
as
\begin{equation}
        H(X) = \cos X ( H_0 + U \tan X) \;.
\end{equation}
Let us also parameterize its eigenvector as $\psi = (\cos \theta,
\sin \theta)$.
Then, by defining the vector $\bbox{h}(X)\equiv (2H_{12}, H_{22} -
H_{11})$, it is straightforward to show that the eigenvector equation
can be written as
\begin{equation}
      (2 H_{12}, H_{22} - H_{11}) \cdot (\cos 2\theta, \sin 2\theta) = 0 \;,
\end{equation}
so that the eigenvectors are determined solely by the vector
$\bbox{h}(X)$. Moreover, the angle between $\psi (X+\delta X)$ and
$\psi (X)$ is half the angle $\alpha$ between $\bbox{h}
(X+\delta X)$ and $\bbox{h} (X)$, that is $ |S|=\cos \alpha/2 $.
Now the problem is reduced to finding the distribution of
$\alpha$. Let us set $X=0$ and introduce
\begin{eqnarray}
\bbox{h}(X=0) &
\equiv &
       \bbox{h}_0 = \big(2[H_0]_{12}, [H_0]_{22}-[H_0]_{11}\big)
\nonumber \\
\bbox{h}(\delta X) & \equiv &
       \bbox{h}_t = \big(2[H_0]_{12}, [H_0]_{22}-[H_0]_{11}\big) +
\nonumber \\
                   &        &
       \hspace{2PC} (2U_{12}, U_{22}-U_{11})t \;,
\end{eqnarray}
with $t = \tan \delta X$. The usefulness of this geometrical
construction becomes clear now. We use the fact that $\bbox{h}_t =
\bbox{h}_0 + \bbox{u}t$ and
\begin{equation}
  \bbox{h}_t \cdot \bbox{h}_0 = |\bbox{h}_t| | \bbox{h}_0| \cos\alpha \;,
\end{equation}
where the vectors $\bbox{h}_t$ and $\bbox{h}_0$ can be expressed in
terms of $|h_0|$, $|u|$, $\theta_u$, and $\alpha$.
The integration of $P(\bbox{h}_0, \bbox{h}_t)$ over $h_0, u$ and $\theta_u$,
readily gives $P(\alpha)$.

The distribution $P(\bbox{h}_0, \bbox{h}_t)$ is written in terms of $H_0$ and
$U$ matrix elements, Eq.~(\ref{eq:P(H)}), as
\begin{equation}
P(\bbox{h}_0, \bbox{h}_t)  = t^2 P\!\left(\bbox{h}_0,
                        \bbox{u}=\frac{1}{t}[\bbox{h}_t - \bbox{h}_0]\right)\;.
\end{equation}
The integration over the vectors $\bbox{h}_0$ and $\bbox{h}_t$,
keeping  their relative orientation fixed, yields
\begin{equation}
P(\alpha) = \frac{2t^2}{\pi} \int_0^{\pi/2} \! d\phi
            \frac{\cos \phi \sin \phi}
            {(1+t^2\cos^2\phi-2\cos\phi \sin\phi \cos\alpha)^2}  \; .
\end{equation}
The latter is expressed in a closed form as
\begin{equation}
P(\alpha) = \frac{t^2}{\pi \left( \sin^2 \alpha +t^2 \right)}
     \big[ 1 + G(\alpha)\big]  \; ,
\end{equation}
with
\begin{equation}
G(\alpha)= \frac{\cos \alpha} {\sqrt{\sin^2 \alpha +t^2}}
           \left[ \frac{\pi}{2} +
                  \tan^{-1} \left( \frac{\cos \alpha}{\sqrt{\sin^2 \alpha +t^2}}
                                       \right)
           \right] \; .
\end{equation}
Since we are not really interested in $P(\alpha)$, but rather in
$P(|\widetilde{S}|)$ instead, a last change of variables is needed
to arrive at the main result of this section
\begin{equation}
\label{eq:PS2x2}
P(|\widetilde{S}|) = 2 \frac{ P[\alpha = 2 \, \cos^{-1} |\widetilde{S}|]}
{\sqrt{1-|\widetilde{S}|^2}}       \; .
\end{equation}

In particular, the probability density of having a null
overlap is non-vanishing and is given by
\begin{equation}
P(|\widetilde{S}|=0) = \frac{2}{\pi}\left( 1-\frac{\delta X}{\tan \delta X} \right) \; .
\end{equation}

\begin{figure}
\setlength{\unitlength}{1mm}
\begin{picture}(90,70)
\put( 25,  -27){\epsfxsize=65mm\epsfbox[200 208 600 600]{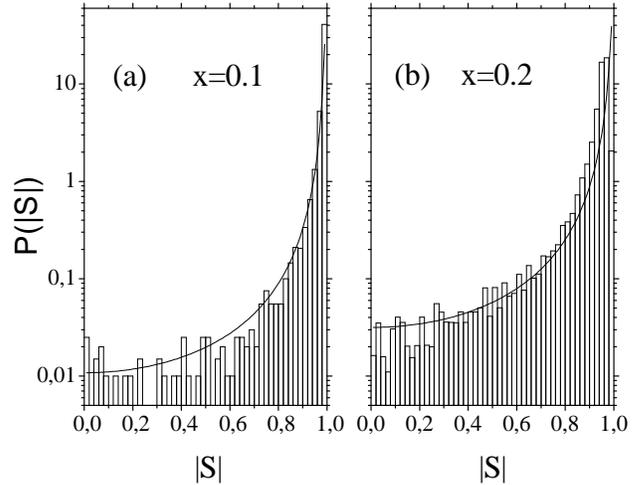}}
\end{picture}
\caption{Log-normal plot of the distribution of overlaps $P(|S|)$.
We compare the analytical prediction based on the case $N=2$ 
with numerical simulations over $10^5$ pairs $(H_0,U)$ (histograms), 
for $N=100$. The relative deformation is (a) $x=0.1$ and (b) $x=0.2$.}
\label{fig:hist2x2}
\end{figure}

Comparison between the analytic form of $P(K\cdot
|\widetilde{S}|)$, Eq.~(\ref{eq:PS2x2}) [cf.\ also
Eq.~(\ref{eq:2x2a})], and our numerical study of the exact
distribution of overlaps $P(|S|)$ is displayed in
Fig.~\ref{fig:hist2x2}. It is evident that the 2-level picture
reproduces
 the tails of the distributions $P(|S|)$ for large $N$
remarkably well,  provided we use the same relative deformation
$x$. Note that for $N=2$ and $\beta=1$, one has
$X^\star=\sqrt{\pi}$, differing from the large $N$ limit
given by Eq.\ (\ref{eq:defXstar}), see for instance Ref.
\onlinecite{Bohr69}. The correction due to $K_N$, a rescaling of
the horizontal axis, is negligible in this case (from
Fig.~\ref{fig:lastlevel} one finds that $K\approx 0.98$)  and was
not included.

\begin{figure}
\setlength{\unitlength}{1mm}
\begin{picture}(100,75)
\put( 25, -3){\epsfxsize=55mm\epsfbox[200 208 600 600]{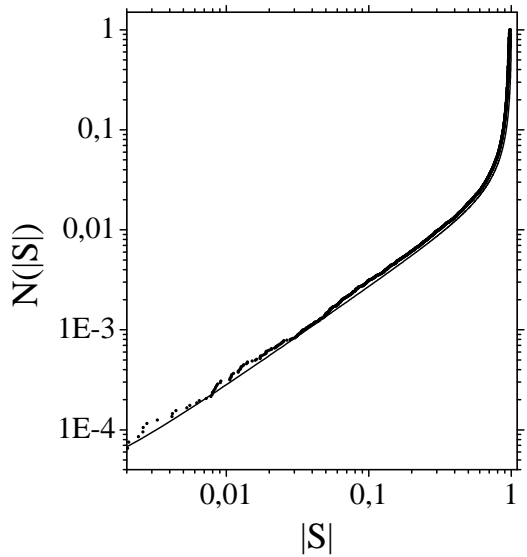}}
\end{picture}
\caption{Log-log plot of the cumulative distribution of overlaps
$N(|S|)=\int_0^{|S|} d|S|\,P(|S|)$. We compare (i) the analytical
prediction based on the case $N=2$  and (ii) numerical simulations
over $10^5$ pairs $(H_0,U)$ (dots), for $N=50$. In both cases the
relative deformation is $x=0.2$.}
\label{fig:hist2x2log-log}
\end{figure}

The agreement between the output of our $2 \times 2$ model
and the exact diagonalization becomes even more evident
by analyzing the cumulative overlap distribution in a log-log
plot, as is seen from Fig. \ref{fig:hist2x2log-log}

\noindent

%-------------------------------------------------------------------------
\section{Concluding remarks}
\label{sec:conclusions}
%-------------------------------------------------------------------------

We have studied the orthogonality catastrophe due to a parametric
change of the single-particle of an ergodic system. The
Hamiltonian was modelled by a suitable random matrix ensemble. We
show that the average overlap between the original and the
parametrically modified many-body ground states, taken as Slater
determinants, decreases like  $n^{-\beta x^2/4}$, where $n$ is the
number of electrons in the system
and $x$ is the deformation measured in units of the typical
distance between anticrossings. We have also shown that the
fluctuations of $\log |S|$ are enormous. To account for the latter
in the orthogonal case ($\beta=1$), we have put forward a simple
$2 \times 2$ matrix model and employed it to obtain a prediction
for $P(|S|)$ for $|S| \ll 1$, in good agreement with our numerical
analysis. In the unitary case ($\beta=2$) the fluctuations are
smaller, but still quite significant. Here also it was shown
\cite{Yuval} that the $2 \times 2$ model works well, but no simple
analytical expression is avaliable.

This model study constitutes a first step towards understanding
the relevance of the orthogonality catastrophe for ballistic
ergodic systems. One improvement to the theory presented in
Section \ref{sec:results} should arise from the fact that for a
generic dynamical system the ``perturbation" matrix, $\delta
H_{ij}$, will in general have a finite bandwidth $b$. How are our
results changed in this case? There is extensive literature
dealing with the matrix elements properties of low-dimensional
dynamical systems. In particular, for the case of a
two-dimensional chaotic ballistic system, the bandwidth for a
``generic" perturbation was estimated \cite{Feingold86,Cohen} to
be
\begin{equation}
   b = \frac{\hbar^2}{\tau \Delta} \; ,
\end{equation}
where $\tau$ is a time needed for a particle to traverse
the system, i.e.,
\begin{equation}
   \tau \approx \frac{L}{v_F} \; .
\end{equation}
Here $L$ is the linear dimension of the system and $v_F$ is the
Fermi velocity. Relating the 2D mean level spacing to the system's
size, $\Delta = 2 \pi \hbar^2 / (m L^2) $, and using $\varepsilon_F 
\approx n \Delta $, we obtain a simple relation between
the bandwidth and the number of electrons, namely,  $ b \approx \sqrt{n}$.
In line with the present study, we assume that within the band
$|j-i|<b$, the fluctuations of $\delta H_{ij}$ are statistically
well described by RMT. Accordingly, the average of $\log |S|$,
obtained in Eqs. (\ref{eq:s_average}) , (\ref{eq:s_average2}),
becomes
\begin{equation}
\label{eq:band}
   \left< \log |S| \right> \approx - \frac{\beta}{4} x^2 \log \sqrt{n} \; .
\end{equation}
We stress that in order to write Eq.~(\ref{eq:band}), two
important assumptions have been made. (i) The Hamiltonian  $H(X)$
must be fully chaotic. \cite{Feingold86} (ii) The perturbation
must be generic in the sense defined in Ref.~\onlinecite{Cohen}.
If such conditions are not met, the statistical distribution of
$\delta H_{ij}$ could strongly depend on  $|j-i|$, for $|j-i|<b$.

The study of the overlap between of many-body states subjected to
a perturbation has also been  a traditional subject of
investigation in nuclear physics problems  \cite{Ring}. More
recently there is a renewed interest in examining  overlaps among
excited states \cite{Kusnezov96}. This study is much more involved
than ours, since the  states considered  are superpositions of
Slater determinants, rather than a single one as studied here. A
full theory for such a case is still lacking; available numerical
evidence (evidently for small systems) indicates that the scaling
of the overlaps  with $n$ follows a Gaussian, a fact  which is yet
to be  explained.

%-------------------------------------------------------------------------
\acknowledgements
We thank D. Kusnezov, I. V. Lerner, E. Mucciolo, M. Saraceno,
and  E. Vergini for helpful discussions. This work
was supported by FAPERJ, CNPq and PRONEX (Brazil), and by the
U.S.-Israel Binational Science Foundation, the DIP Foundation, the
Minerva foundation and The Israel Science Foundation  founded by
the Israel Academy of Sciences-Center of Excellence Program
(Israel).

%-------------------------------------------------------------------------
\appendix
%-------------------------------------------------------------------------

\section{A perturbation series for
         {\mbox{\boldmath $\log S$}}}
\label{app:perturbation}

In Section \ref{sec:OC} we showed that the overlap of the ground
states $S$ can be written as $S =  \det A^{\rm oo}$, with the matrix
elements of $A$ defined by Eq.~(\ref{eq:defA}).
This is done for the case where time-reversal symmetry is present
(absent), namely  the orthogonal (unitary) symmetry.
The notation ``oo" stands for ``occupied-occupied'' states (in this
Appendix we shall also use ``oe" for ``occupied-empty'').
Within the Rayleigh-Schr\"odinger perturbation theory the overlap
matrix is expanded as
\begin{equation}
A = \mbox{$\openone$} + \epsilon B + \epsilon^2 C + \epsilon^3 D + \ldots \;.
\label{expansion}
\end{equation}
Since $AA^\dagger=\mbox{$\openone$}$, i.e., using the fact that this series is
normalized, it is implied that
\begin{eqnarray}
 B + B^\dagger                              & = & 0                       \nonumber\\
 C + C^\dagger + B B^\dagger                & = & 0 \label{orthogonality} \nonumber\\
 D + D^\dagger + C B^\dagger + B C^\dagger  & = & 0 \;,
\end{eqnarray}
etc..
In particular, one knows from perturbation theory that the first order
correction $B$ is an antihermitian matrix satisfying $B_{ii}=0$ and
\begin{equation}
\epsilon B_{ij} = \frac{\delta H_{ij}}{E_j-E_i}, \qquad \mbox{for} \qquad
            j>i \;.
\end{equation}
To obtain a perturbative series for $\log |S|$ we use Eq.~(\ref{expansion})
and the identity $\log \det  = \mbox{tr} \log $ to write
\begin{equation}
\label{eq:series}
 \log \det A^{\rm oo}   = \mbox{tr}
                        \log \left( \mbox{$\openone$} +
                                  \epsilon   B^{\rm oo} +
                          \epsilon^2 C^{\rm oo} +
                          \epsilon^3 D^{\rm oo} +
               \ldots \right )                         \;.
\end{equation}
By expanding the logarithm and regrouping the terms in the sum
order by order in $\epsilon$, we obtain
\begin{equation}
 \log \det A^{\rm oo}  = \mbox{tr} \left[ \epsilon    B^{\rm oo} +
                                        \epsilon^2 (C^{\rm oo} -
                                          \frac{1}{2}      {B^{\rm oo}}^2) +
                   \ldots \right]    \; .
\end{equation}
The linear term in $\epsilon$ vanishes since $B_{ii}=0$.
Using the first two equations in (\ref{orthogonality}) one arrives at
\begin{equation}
 \log \det A^{\rm oo}  =
       - \frac{1}{2} \epsilon^2 \mbox{tr}
         \left( B^{\rm oe} {B^{\rm oe}}^\dagger -
                C^{\rm oo} + {C^{\rm oo}}^\dagger \right) +
                      {\cal O}(\epsilon^3) \;.
\end{equation}
The contribution from $C^{\rm oo}-{C^{\rm oo}}^\dagger$ being purely
imaginary corresponds to a phase in $\det A^{\rm oo}$, so that
\begin{equation}
 \log | \det A^{\rm oo} \,|  =
       - \frac{1}{2} \epsilon^2 \mbox{tr}
         B^{\rm oe} {B^{\rm oe}}^\dagger + {\cal O}(\epsilon^3) \;.
\end{equation}
If we average over the parametric random matrix ensemble,
Eq.~(\ref{eq:variance}), the third order terms disappear:
\begin{equation}
      \langle \log |\det A^{\rm oo} \,| \rangle =
     - \frac{1}{2} \epsilon^2 \mbox{tr} \langle B^{\rm oe} {B^{\rm oe}}^\dagger \rangle +
                      {\cal O}(\epsilon^4)           \; ,
\end{equation}
or, equivalently,
\begin{equation}
      \langle \log |\det A^{\rm oo} \,| \rangle =
      - \frac{1}{2} \sum_{ij} \left<
       \frac{|\delta H_{ij}|^2}{(E_j-E_i)^2}   \right>  +
                      {\cal O}\left[(\delta H)^4 \right]\; ,
\end{equation}
where $j$ and $i$ run over the occupied and empty states,
respectively.

This calculation also shows that
\begin{equation}
  \exp \langle \log |\det A^{\rm oo} \,| \rangle -
       \langle      |\det A^{\rm oo} \,| \rangle  =
             {\cal O} (\epsilon^4) \;.
\end{equation}

%-------------------------------------------------------------------------

\section{Grand canonical averages}
\label{app:GC}

In this appendix we study a slightly different question,
complementary to the ones we have addressed so far. It is related
to the averaging procedure. Suppose we have an ensemble of
systems, each of which having a given number  of electrons, $N$,
which is 
kept fixed as the parameter $X$ is varied. Let us assume,
though, that the {\it preparation} of this ensemble of systems is
performed {\it grand canonically}, i.e., by attaching each sample
to a weakly coupled particle reservoir at a given chemical
potential $\mu$, equilibrating, and then removing this coupling.
The number of electrons in each system may vary, but is kept fixed
during the ``measurement" (i.e., varying $X$). This procedure has
been defined in Refs.\onlinecite{Kamenev93,Kamenev94} as {\it
grand-canonical---canonical}. Under these conditions  we employ a
grand-canonical averaging scheme to calculate $\langle \log |S|
\rangle$. Here we expect that large fluctuations due to avoided
crossings will be suppressed, since the statistical weight of an
avoided crossing needs to be modified from $P(s)$ (canonical) to
$sP(s)$ (grand canonical) \cite{Shklovskii82}. This reflects the
fact that the probability to  place the chemical potential in a
given gap (between the last occupied and the first vacant level)
is proportional to the size of the gap.

\begin{figure}
\setlength{\unitlength}{1mm}
\begin{picture}(70,65)
\put( 7, -39){\epsfxsize=100mm\epsfbox[79 158 682 652]{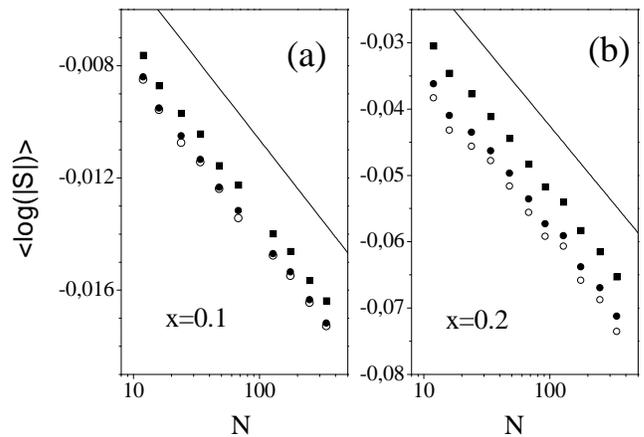}}
\end{picture}
\caption{The average $\log |S(x)|$ as a function of $N$.
         The number of realizations for each $N$ is $M=10^4$
         and (a) $x=0.1$ and (b) $x=0.2$.
         Open dots stand for the exact $\langle \log |S| \rangle$,
         filled dots for $\langle I_{\mbox{\scriptsize norm}}\rangle$,
         squares for $\langle I \rangle$, whereas the solid lines
         represent $\langle \widetilde{I} \rangle$. See Section 
         \ref{sec:results} for the definitions.}
\label{fig:grandcanonical}
\end{figure}

The results of our simulations using the grand canonical ensemble 
are summarized by Fig.~\ref{fig:grandcanonical}. There, as in Fig.
\ref{fig:canon-average}, we show $\langle \log |S| \rangle$ and its 
different approximation schemes (explained in Section \ref{sec:results})
as a function of $N$.

By comparing the results of Fig. \ref{fig:canon-average} with the ones 
shown above we observe that the values of $\langle \log |S| \rangle$,
as well as $\langle I_{\mbox{\scriptsize norm}}\rangle$ and 
$\langle I \rangle$, obtained by the canonical simulation are 
systematically reduced with respect to the ones obtained 
using the grand canonical averaging procedure.
This is in line with the reasoning that the grand canonical averaging
suppresses large fluctuations corresponding to small gaps. Thus it eliminates 
the long small $|S|$ tails of $P(|S|)$ characteristic of the canonical 
ensemble for $x \ll 1$.
We also observe that both ensemble averaging procedures tend to render 
similar results as the value of $x$ is increased. (For this reason we 
refrain from showing the grand canonical averages corresponding to $x=0.5$ 
and $x=1.0$ as depicted in Fig. \ref{fig:canon-average}.)

\begin{figure}
\setlength{\unitlength}{1mm}
\begin{picture}(70,70)
\put( 10, -15){\epsfxsize=80mm\epsfbox[79 158 682 652]{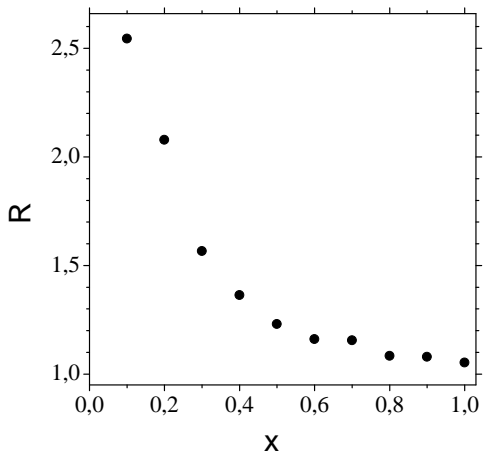}}
\end{picture}
\caption{The ratio between canonical and grand canonical standard
         deviation $\delta\log |S(x)|$ as a function of $x$ for
         a fixed $N=50$ and $M=10^4$.}
\label{fig:ratio-variance}
\end{figure}

Figure \ref{fig:ratio-variance} quantifies the discussion presented in the
foregoing paragraph. The object of study is the standard deviation
$\delta \log|S| = [\langle(\log|S|)^2\rangle - \langle \log |S| \rangle^2]^{1/2}$.
For a fixed matrix size $N=50$ and $M=10^4$
realizations we compute the ratio $R$ between the canonical and the grand 
canonical $\delta\log |S|$ as a function of $x$. As expected 
$R(x) > 1$ for all investigated values of $x$. Furthermore, as we increase 
$x$ the occurrence of small gaps becomes less important and the canonical 
fluctuations get closer to the grand canonical ones.

%-------------------------------------------------------------------------

\section{Accuracy of the approximations for the average of single particle
         overlaps}
\label{app:FOPT}

\begin{figure}
\setlength{\unitlength}{1mm}
\begin{picture}(70,70) 
\put( 20,  25){\epsfxsize=65mm\epsfbox[200 208 600 600]{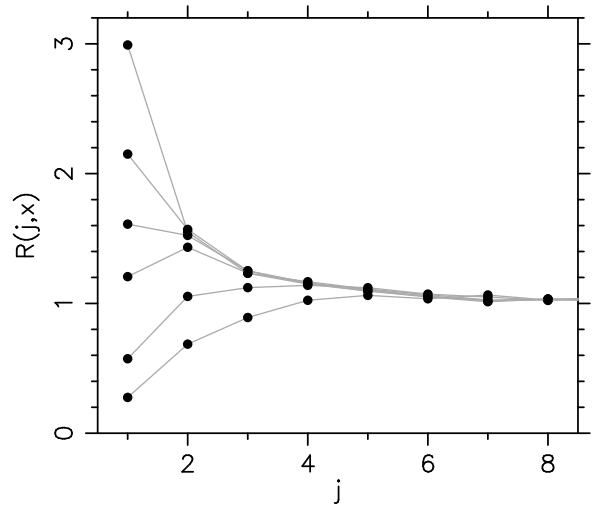}}
\end{picture}
\caption{Test of first order perturbation theory plus average spectrum.
We plot the ratio $R$, the quotient ``exact/approximate'' (see text),
as a function of the energy difference $j$ for
different values of the parametrical distance $x=\delta X/X^\star$
($x = 0.2, 0.3, 0.4, 0.5, 0.75, 1.0$; the curves with
increasing $x$ correspond to the ones with decreasing values for
the first abscissa point $j = 1$). Each curve was calculated
by averaging over 10000 pairs $(H_0,U)$ with  $N=100$.}
\label{fig:FOPT}
\end{figure}

Here we exhibit a test of the combination of first order perturbation
theory and an average spectrum to account for averages of single particle
overlaps. We want to compare the average
\begin{equation}
\label{eq:app-1}
\langle | A_{N/2,N/2+j} |^2\rangle =
  \left< |\langle \psi_{N/2}(X) | \psi_{N/2+j}(X+\delta X) \rangle|^2 \right>
\end{equation}
with the approximation discussed in Section~\ref{sec:results}
\begin{equation}
\label{eq:app-2}
\langle|\widetilde{A}_{N/2,N/2+j}|^2\rangle =
\frac{\beta x^2 \Delta^2}
           {2 \big\langle \varepsilon_{N/2+j} - \varepsilon_{N/2} \big\rangle^2} \; .
\end{equation}
The quantity plotted in Fig. \ref{fig:FOPT} is the ratio
$R = \langle |A_{N/2,N/2+j}|^2\rangle/ \langle |\widetilde{A}_{N/2,N/2+j}|^2
\rangle$.
For $j$ small as compared to $N/2$ the average semi-circle spectrum can
be approximated by a ``picket fence'' with spacing $\Delta$, meaning that
$\langle \varepsilon_{N/2+j} - \varepsilon_{N/2} \rangle^2 = j^2\Delta^2$.
The ratio $R$ is then
\begin{equation}
R(j,x)=
      \frac {2 j^2}{\beta x^2}
\left< |\langle \psi_{N/2}(X) | \psi_{N/2+j}(X+\delta X) \rangle|^2 \right>
              \, .
\end{equation}
The figure clearly displays the following features.
For $x$ fixed, the ratio $R \rightarrow 1$
for large values of $j$. When $x<1$ the approximation works
well except for the few states with, say, $j \le 5$.
In the case of $j=1$, even if $x \ll 1$, perturbation
theory breaks down due to very narrow avoided crossings.

%--------------------------------------------------------------------------

\end{multicols}


\begin{thebibliography}{99}

\bibitem{Anderson67}
   P. W. Anderson,
        Phys. Rev. Lett. {\bf 18}, 1049 (1967).

\bibitem{review}
  L. P. Kouwenhoven, C. M. Marcus, P. L.~McEuen, S.~Tarucha,
  R. M.~Westervelt, and N. S.~Wingreen in {\it Mesoscopic Electron
  Transport}, edited by L. L. Sohn, L. P. Kouwenhoven, and
  G.~Sch{\"o}n (Kluwer, Dordrecht, 1997)

\bibitem{Alhassid00}
   Y. Alhassid,
        Rev. Mod. Phys. {\bf 72}, 895 (2000).

\bibitem{Aleiner01}
   I. L. Aleiner, P. W. Brower  and L. I. Glazman,
        cond-mat/0103008

\bibitem{Ben-Jacob85}
   E. Ben-Jacob and Y. Gefen,
        Phys. Lett. {\bf 108A}, 289 (1985);
   E. Ben-Jacob, Y. Gefen, K. Mullen, and Z. Schuss,
        Phys. Rev. B {\bf 37}, 7400 (1988).

\bibitem{Mehta91}
   M. L. Mehta,
        {\sl Random Matrices}, 2{\sl nd} Edition
        (Academic Press, New York, 1991).

\bibitem{Guhr98}
   T. Guhr, A. M\"uller-Groeling, and H. A. Weidenm\"uller,
        Phys. Rep. {\bf 299}, 190 (1998).

\bibitem{Kamenev95}
   A. Kamenev and Y. Gefen,
        unpublished (1995).

\bibitem{Blanter97}
   Ya. M. Blanter, A. D. Mirlin, and B. A. Muzykantskii,
        Phys. Rev. Lett. {\bf 78}, 2449 (1997).

\bibitem{Kurland00}
   I. L. Kurland, I. L. Aleiner, and B. L. Altshuler,
        cond-mat/0004205.

\bibitem{Koopmans}
   T. Koopmans,
        Physica {\bf 1}, 104 (1934).

\bibitem{Walker99}
   P. N. Walker, G. Montambaux, and Y. Gefen,
        Phys. Rev. B {\bf 60} 2541 (1999).

\bibitem{Vallejos98}
   R. O. Vallejos, C. H. Lewenkopf, and E. R. Mucciolo,
        Phys. Rev. Lett. {\bf 81}, 677 (1998).

\bibitem{coulomb-blockade}
   C. W. J. Beenakker,
        Phys. Rev. B {\bf 44}, 1646 (1991).

\bibitem{Chen92}
   Y. Chen and J. Kroha,
        Phys. Rev. B {\bf 46}, 1332 (1992);
   I. L. Aleiner and K. A. Matveev,
        Phys. Rev. Lett. {\bf 80}, 814 (1998);
   I. E. Smolyarenko and B. D. Simons,
        unpublished.

\bibitem{Simons}
   B. L. Altshuler and  B. D. Simons in
   {\it Mesoscopic Quantum Physics},
   edited by E. Akkermans, G. Montambaux,
   J.-L. Pichard and J. Zinn-Justin (North-Holland, Amsterdam)
   p.1 .

\bibitem{Alhassid01}
   Y. Alhassid and Y. Gefen,
        cond-mat/0101461.

\bibitem{Alhassid2001b}
   Y. Alhassid and Y. Gefen,
        to be published.

\bibitem{Wilkinson89}
   M. Wilkinson,
        J. Phys. A {\bf 22}, 2795 (1989);
   Y. Alhassid and H. Attias,
       Phys. Rev. B {\bf 54}, 2696 (1996).

\bibitem{Szafer93}
   A. Szafer and B. L. Altshuler,
        Phys. Rev. Lett. {\bf 70}, 587 (1993).

\bibitem{Simons93}
   B. D. Simons and B. L. Altshuler,
        Phys. Rev. Lett. {\bf 70}, 4063 (1993).

\bibitem{Shklovskii82}
   B.I. Shklovskii,
        Pis'ma Zh. Eksp. Teor. Fiz. {\bf 36}, 287 (1982)
        [JETP Lett. {\bf 36}, 352 (1982)].

\bibitem{Kamenev94}
   A. Kamenev and Y. Gefen,
        Phys. Rev. B {\bf 49}, 14474 (1994);
   A. Kamenev and Y. Gefen in {\it Quantum Dynamics of
Submicron Structures}, edited by H. A. Cerdeira {\it et. al.}
(Kluwer Academic Publ., The Netherlands, 1995) pp. 81.

\bibitem{Yuval}
   Y. Gefen, R. Berkovits, I. V. Lerner, and B. L. Alshuler,
        to be published.

\bibitem{Bohr69}
   A. Bohr and B. R. Mottelson,
        {\sl Nuclear Structure}, vol. 1 (W. A. Benjamin, New York, 1969).

\bibitem{Kamenev93}
   A. Kamenev and Y. Gefen,
        Phys. Rev. Lett. {\bf 70}, 1976 (1993).

\bibitem{Feingold86}
   M. Feingold and A. Peres,
      Phys. Rev. A {\bf 34}, 591 (1986).

\bibitem{Cohen}
   D. Cohen, F. M. Izrailev, and T. Kottos,
      Phys. Rev. Lett. {\bf 84}, 2052 (2000);
   A. Barnett, D. Cohen, and E. J. Heller,
      Phys. Rev. Lett. {\bf 85}, 1412 (2000);
   D. Cohen,
      Ann. Phys. {\bf 283}, 175 (2000);
   D. Cohen and T. Kottos,
      Phys. Rev. E. {\bf 63}, 36203 (2001).

\bibitem{Ring}
   P. Ring and P. Schuck,
      {\it The Nuclear Many-Body Problem},
      (Springer, New York, 1980). 

\bibitem{Kusnezov96}
   D. Kusnezov, B. A. Brown, and V. Zelevinsky,
      Phys. Lett. B {\bf 385}, 5 (1996).


\end{thebibliography}
\end{document}